\documentclass[runningheads,anonymous]{llncs}
\usepackage{graphicx}
\usepackage{algorithm}
\usepackage{algpseudocode}
\usepackage{parskip}
\usepackage[table,xcdraw]{xcolor}
\usepackage{booktabs}
\usepackage{multirow}
\usepackage{lscape}
\usepackage{array}
\usepackage{pifont}
\usepackage{makecell}

\usepackage[table,xcdraw]{xcolor}

\begin{document}
%
\title{ISSF: The Intelligent Security Service Framework for Cloud-Native Operation}
\titlerunning{The Intelligent Security Service Framework for Cloud-Native Operation}
%
%


\author{Yikuan Yan\inst{1} \and
Keman Huang\inst{2,*} \and
Michael Siegel\inst{3}}

\authorrunning{Y Yan et al.}

\institute{School of Information, Renmin University of China, Beijing, China
\email{yanyikuan@ruc.edu.cn} \and
School of Information, Renmin University of China, Beijing, China;
Cybersecurity at MIT Sloan, MIT, Cambridge, Massachusetts, USA\\
\email{keman@ruc.edu.cn} \and 
Cybersecurity at MIT Sloan, MIT, Cambridge, Massachusetts, USA\\
\email{msiegel@mit.edu}
}

\maketitle              
%

\begin{abstract}

The growing system complexity from microservice architectures and the bilateral enhancement of artificial intelligence (AI) for both attackers and defenders presents increasing security challenges for cloud-native operations. In particular, cloud-native operators require a holistic view of the dynamic security posture for the cloud-native environment from a defense aspect. Additionally, both attackers and defenders can adopt advanced AI technologies. This makes the dynamic interaction and benchmark among different intelligent offense and defense strategies more crucial. Hence, following the multi-agent deep reinforcement learning (RL) paradigm, this research develops an agent-based intelligent security service framework (ISSF) for cloud-native operation. It includes a dynamic access graph model to represent the cloud-native environment and an action model to represent offense and defense actions. Then we develop an approach to enable the training, publishing, and evaluating of intelligent security services using diverse deep RL algorithms and training strategies, facilitating their systematic development and benchmark. The experiments demonstrate that our framework can sufficiently model the security posture of a cloud-native system for defenders, effectively develop and quantitatively benchmark different services for both attackers and defenders and guide further service optimization.

\end{abstract}
\section{Introduction}

The cloud-native and microservice approach has been increasingly adopted for cloud services design, development, and deployment\cite{carrusca2020microservices,he2019re}. As a modern software development and deployment methodology, it designs applications as loosely coupled microservices interact through API endpoints and uses container technology to provide lightweight runtime environments. While this approach significantly simplifies the updating, scheduling, and scaling of cloud-native systems, the security threat has arisen.


First, the microservice-based unbundling and the container-based lightweight virtualization technology cause an increasingly difficult-to-control attack surface for defense \cite{yu2019survey}. cloud-native defenders can easily get lost in complex configurations and interactions without a precise understanding of the cyber threat situations. A recent direction to bridge this gap is to model the cloud-native system from an attack graph aspect to optimize cyber defense strategies in specific security scenarios, like Man-in-the-Middle attack or container escape \cite{zdun2023microservice,jin2019dseom}, rather than from a defense operation aspect. This motivates us to develop a model for cloud operators to understand the dynamic surface of the cloud-native environment.

Second, the increasingly advanced AIs have significantly powered not only the cyber defenders but also attackers, demonstrated as increasing autonomous intelligent cyber defense and offense\cite{vyas2023automated} services. Autonomous cyber operations (ACOs), typically like moving target defense (MTD), have been developed as an effective defense strategy \cite{standen2021cyborg,jin2019dseom}. While some recent studies optimize the defense strategies for cloud-native defenders, these strategies are designed with a specific deployment setting, making it challenging to compare different strategies in a systematic way\cite{ma2023mutation}. Additionally, cyber attackers can also adopt similar, if not more advanced, AI techniques. For example, in a cloud-native environment, attackers can set up a cloud-native environment to pre-train an intelligent offense service\cite{huang2018systematically,pearlson2022design} and then use it to guide the cyber offense. While existing studies take the perspective from either the attacker \cite{ma2023mutation} or the defender \cite{jin2019dseom}, it is essential to consider them as a whole, especially their dynamic interactions, to support effective cloud-native security operation.

Hence, this research proposes the intelligent security service framework, named \textit{ISSF}, to investigate intelligent security services for both attackers and defenders within the cloud-native environment. First, ISSF establishes a dynamic access graph model from a defense perspective to represent the security surface for a cloud-native environment. Each attacker or defender is defined as an agent-based intelligent security service that can undertake offense or defense actions, including three offense actions (\textit{local attack, remote attack and connect}) and three defense actions (\textit{scan, restore and remediate}). These can be further combined to form complex tactics. Second, ISSF provides a flexible and extensible approach which can \textit{train} offense or defense intelligent security services using diverse deep reinforcement learning algorithms from scratch or \textit{finetune} from a pre-trained security service, and then \textit{publish} it to the \textit{security service pool}. We further develop an ELO rating-based approach to quantitatively \textit{evaluate} the strength of different offense or defense intelligent security services. Finally, we propose a three-service-chain cloud-native case to verify our framework. While the experiments demonstrate the effectiveness of our framework, the preliminary results reveal that training a security service using more advanced and diverse adversaries could achieve a better performance, suggesting a promising direction for security service optimization. Overall, our framework contributes:
 
\begin{itemize}
      \item An agent-based intelligent security service model includes a dynamic access graph model to represent security situations and an action model to represent the defense and offense actions.
      \item A flexible and extensible approach that can train, publish and evaluate intelligent security services in a systematic and quantitative way.
      \item A case designed to verify the effectiveness of our framework and provide empirical evidence to guide further security service optimization.
\end{itemize}

The rest of this paper is structured as follows. Section 2 discusses the related work to position this research. Section 3 describes the agent-based intelligent security service model. We detail our approach for training, publishing, and evaluating security services in Section 4. The experiment results are reported in Section 5. Section 6 concludes this paper and discusses the limitations.

\section{Related Work}

\subsubsection{cloud-native Security Modeling.} The cloud-native security has witnessed extensive research efforts in various aspects. The security modeling for cloud-native environments has become increasingly prevalent\cite{jin2019dseom,ma2023mutation,zdun2023microservice}. For example, Jin et al. develop a holistic attack graph, which depicts attack scenarios in container-based cloud environments \cite{jin2019dseom}. Ma et al. devise a threat model and propose a mutation-enabled proactive defense strategy specifically targeting Man-in-the-Middle attacks \cite{ma2023mutation}. U. Zdun et al. employ multiple sets of metrics in their microservice system model and develop a microservice architectural design decisions (ADD) model for security strategies\cite{zdun2023microservice}. However, most of these existing models and strategies are developed for specific offense scenarios and model the cloud-native environment from the attack graph aspect. There is a lack of effective cloud-native environment models to support defenders in understanding the security situation from the aspect of dynamic interactions among the loosely coupled microservices.

\subsubsection{Intelligent Cybersecurity Operations and Optimizations.} Deep reinforcement learning has been increasingly adopted for intelligent cybersecurity operations in recent years \cite{standen2021cyborg}. One typical effort is the Autonomous Cyber Operations (ACO), wherein analysis and decision-making processes can be autonomously optimized and performed to safeguard computer systems and network environments\cite{vyas2023automated}. ACO Gyms serving as cyber system environments that enable the deployment of autonomous red and blue team agents have been developed, including CyberBattleSim\cite{msft:cyberbattlesim} and CybORG\cite{standen2021cyborg}. Research efforts further extended their capabilities through component and functional extensions\cite{walter2021incorporating} and algorithmic development\cite{applebaum2022bridging}. However, they overlook the fact that both attackers and defenders can be powered by AI. While the operation strategies are optimized from the aspect of either attacker\cite{ma2023mutation} or defender\cite{jin2019dseom}, without considering attackers and defenders as a whole, it is challenging to evaluate their true effectiveness and further optimize the offense or defense strategies.

\section{The Agent-based Intelligent Security Service Model}

Based on the multi-agent deep reinforcement learning (MADRL) paradigm, we propose the intelligent security service model for cloud-native environments, including a dynamic access graph representing the dynamic surface of a cloud-native system and the action model representing the security operation for attackers and defenders. 

\subsection{Dynamic Access Graph Model for Cloud-Native System}

A cloud-native system is a dynamic system where service instances, the microservice application running on a container, can be easily deployed, updated, or destroyed. Logically, these service instances can access each other through API endpoints or being controlled through credentials, forming the dynamic access surface from a security aspect. Hence, we can define a cloud-native environment as a dynamic access graph as follow.

\textbf{Definition 1.} \textit{A cloud-native environment is a directed dynamic access graph \begin{math}DG_{e}=(N,E)\end{math}, where N is a set of nodes representing the service instances in the cloud-native environment, and \begin{math}E \subseteq N \times N\end{math} is a set of edges where each edge is an access path between the ordered pair of nodes \begin{math}\left (N_{i},N_{j}  \right )\end{math}} that \begin{math}N_{i},N_{j}\in N\end{math} and \begin{math}i \neq j \end{math}, using API endpoints or credentials.

Note that a node represents a service instance, which can be a microservice, a container, or a physical machine in the cloud-native environment \cite{jin2019dseom}, and we did not distinguish them in this research from a security situation aspect.

\textbf{Definition 2.} \textit{A node is a quadruplet \begin{math}N=(AV,STAT,CONS,VULN)\end{math}, where asset values $AV$ define the intrinsic value of the digital asset associated with the node and state $STAT$ represent the running state of the node including $running$ or $reimaging$ which can change according to defender's operations. Constraints $CONS$ define the required $connection$ $credentials$ to access the node. Vulnerabilities $VULN$ are the associated $local$ or $remote$ vulnerabilities which can be exploited by the attackers and result in credential or topological information leaks beyond value lost.}

More specifically, a local vulnerability can be exploited only if the target node becomes $owned$ by the attacker. The remote vulnerabilities on a node with $discovered$ state can be exploited by the attacker using an $owned$ node as the attack source. For example, an attacker can exploit a remote vulnerability in $node_B$ with a $discovered$ state from $node_A$ with an $owned$ state. Importantly, once an attacker can exploit a vulnerability in a given node, he/she can collect the associated access information, such as the API endpoints providing topological information to identify additional nodes in the cloud environment, or the credentials to access and control other nodes. Note that, once attackers can access a node using the required $connection$ $credential$, they can $own$ that node. 

\textbf{Definition 3.}  \textit{An edge indicates the access path \begin{math}E=(SOUR,TARG,CONS)\end{math}, where $SOUR$ represents the source node and $TARG$ represents the target node. $CONS$ describes the connections through API endpoints or access credentials.}

Note that given a dynamic access graph $DG_{e}$ for a cloud-native system, attackers and defenders will have different information dynamically on the system, resulting in different \textit{enhanced} subgraphs. This is because these subgraphs represent the attacker's or defender's observation of the system with additional security posture in a given time $t$ based on their security operations. 

\textbf{Definition 4.}  \textit{The observation space for the attacker, $OB_{a,t}$, at time $t$ is an enhanced subgraph of $DG_{e}$, $OB_{a,t}=(N_{a,t},E_{a,t})$ where $N_{a,t} \in N, E_{a,t} \in E$ and each node is further associated with a security state including $discovered$ or $owned$ by the attacker but without detail information of the associated vulnerabilities, each edge represents the order in which each node is discovered by the attacker until time $t$.}

\textbf{Definition 5.}  \textit{The observation space for the defender, $OB_{d,t}$, at time $t$ is an enhanced subgraph of $DG_{e}$, $OB_{d,t}=(N_{d,t},E_{d,t})$ where $N_{d,t} = N, E_{d,t} = E $, while each node is further associated with a security status including $suspicious$ or $norm$ identified by the defender but without detail information of the associated vulnerabilities, each edge represents the access paths among nodes.}

\subsection{Action Model for Offenses and Defenses}

Intuitively, the security situation of a cloud-native system is shaped by the dynamic interactions between attackers and defenders, where they select actions based on their strategy and their observation of the cloud-native environment. Following the reinforcement learning paradigm, we can define the action model to represent the action taken by either an attacker or defender.

\sloppy
\textbf{Definition 6.} \textit{The action model is a quadruplet $RLA_t=(OB_{a|d,t}, SS_{o|d}, ACT_{a|d,t}, REW_{a|d,t})$. At time $t$, the model takes observations from an attacker $OB_{a,t}$ or a defender $OB_{d,t}$ as inputs, uses the embedded security service $SS_{o|d}$, to generate an action $ACT_{a|d,t}$ while taking such an action can generate reward $REW_{a|d,t}$ and then update observations at $t+1$.}

Note that each security service represents an autonomous, intelligent offense service $SS_o$ or defensive service $SS_d$ published through our security service framework, which will be detailed in Section 4. 


\subsubsection{Action Space for Attackers.}

Rather than organizing the attack actions from the detailed tactics like the MITRE Cloud Matrix\footnote{Our action model can cover all the tactics from the MITRE Cloud Matrix through combinations. Please refer to: https://attack.mitre.org/matrices/enterprise/cloud/}, in this research, we define the attacker's actions based on their attack vectors, including exploiting vulnerabilities or using credentials. In particular, for vulnerability exploitation, attackers can take a local attack through a local vulnerability within an $owned$ node or take a remote attack through a remote vulnerability within an $discovered$ node. 


\textbf{Definition 7.1.} \textit{Local Attack. The attacker strategically selects a local vulnerability to launch an attack on a node with $owned$ state. }

\textbf{Definition 7.2.} \textit{Remote Attack. The attacker exploits a remote vulnerability within a node with $discovered$ or $owned$ state. }

Note that the attacker only needs to $discover$ a node associated with a remote vulnerability and then can leverage an $owned$ node as the source to take a remote attack on such a remote vulnerability. It does not require an API endpoint or access credential to enable such an attack. We also suppose that attackers have the exploit toolkit for all the associated vulnerabilities but without the knowledge of whether a specific node is associated with a specific vulnerability. Hence, once the attacker chooses a target and a specific vulnerability, if and only if the selected vulnerability is associated with the selected target, the attack will succeed and the attacker could harvest the outcome, including the leaked API endpoint or connection credential information. The leaked API endpoint information will enable attackers to identify additional nodes within the system, which become $discovered$ for attackers. Furthermore, if attackers can get the credential for a node, they can adopt the credential to connect to that node and increase their privilege, changing its security state to $owned$.

\textbf{Definition 7.3.} \textit{Connect. The attacker utilizes stolen credentials to access and control a node. A successful connection using the matching credentials will compromise the node, subsequently changing its state to $owned$ by attackers.}

These three actions are abstracted, or \textit{meta action}. In practice, they could be instantiated into an attack tactic, and complex tactics can be modeled as the combinations of these three actions. For example, taking a $remote$ $attack$ action on a $discovered$ node to exploit a remote vulnerability to steal the credentials or create an additional account, and then using the $connect$ action to access the service instance and increase the privilege using the malicious credential, can represent attackers' lateral movement from one node to another within a cloud-native environment.

\subsubsection{Action Space for Defenders}

From the defenders' points of view, supposing they have information regarding the API endpoints and credentials, hence, they have comprehensive knowledge of the network's topology\footnote{We exclude the situation where attackers can add malicious nodes into the system because such a node can be considered as a node that an attacker always "owns".}. However, defenders don't have details of associated vulnerabilities or the real-time security state for each node. As we focus on the cloud-native environment, we consider three defensive actions at the node level.

\textbf{Definition 8.1. }\textit{Scan. The defender conducts a comprehensive scan of all nodes within the environment and identifies $suspicious$ nodes with a certain probability.}

After a scanning, defenders can only mark a node as suspicious rather than confirming whether it was attacked or owned by an attacker. Given the imperfect and high false positive for existing threat detection tools \cite{alahmadi202299}, we adopt a probability strategy to reflect such a reality. More specifically, we assign a 5\% false positive rate, which means ignoring 5\% suspicious behavior in $discovered$ or $owned$ nodes or mistaking 5\% normal nodes as suspicious.

Another action for defenders is to restore a node by removing the current instance and then restart a new one, which is the typical practice for the moving target defense strategy \cite{standen2021cyborg,jin2019dseom} in the cloud-native defense. If the node's security state is $owned$ by the attacker, after restoring, the security state will change to $discovered$, rather than $undiscovered$. This is because, in practice, this action also updates the leaked API endpoints, which will access this newly created node to guarantee the smooth running of the cloud-native system. Hence, attackers still have the related topological information to identify it.


\textbf{Definition 8.2. }\textit{Restore. The defender resorts and re-images a selected node to a new one.}

The third action is to remediate the leaked information within a given node by updating the information embedded in it, such as the credentials or API endpoint information. It will nullify the attacker's information collected from previous vulnerability exploitation\footnote{If multiple vulnerabilities within a node are already exploited by an attacker, we will reset only one randomly.}. For example, if the exploitation results in identifying the address for a node, the $remediate$ action, which can be changing the API endpoint for that node in practice, will make the state of that node to $undiscovered$ for attackers.

\textbf{Definition 8.3. }\textit{Remediate. The defender modifies the information embedded in the node, such as the API configurations or credentials, which nullifies the attacker's information collected from related vulnerability exploitation.}

Similarly, defenders can take a combination of these actions to implement some complex defense tactics. For example, for an $owned$ node, if defenders take a $restore$ action, followed by a $remediate$ action on those nodes linked to it, the node will become out of the radar for the attacker.

\subsubsection{Reward}

The reward design is critical for strategy optimizations. The reward for an action depends on its gain and cost.

\textbf{Definition 9. }\textit{Reward. For attackers and defenders, $Reward = Gain-Cost$.}

More specifically, for the $local$ $attack$ and $remote$ $attack$ by exploiting related vulnerabilities, following the previous study \cite{jin2019dseom}, we use the Common Vulnerability Scoring System (CVSS)\footnote{https://nvd.nist.gov/vuln-metrics/cvss} scores to define the reward, where the $Impact Subscore$ measures the gains and the $Exploitability Subscore$ measures the costs for exploiting the vulnerability. For the $connect$ action, the gain is the intrinsic value of the node, and the cost is a fixed value, which will be both predefined.  

Additionally, we consider the offense and defense as a zero-sum battle. Hence, the rewards from a defense action are the same as the related attacks. More specifically, the gains and costs for $Remediate$ vulnerabilities are the same as those of exploiting vulnerabilities. $Restoring$ an $owned$ node will reward the defender with the intrinsic value of that node. However, the $restore$ action has a higher cost than the cost for the $connect$ attack because it requires reimaging the node, which reduces the system availability for a cloud-native system. Finally, there is a small, predefined fixed cost for the $scan$ action, and its gain depends on the number of newly discovered $suspicious$ nodes. In other words, if there are no additional suspicious nodes identified through the $scan$ action, there will be no gain but cost, resulting in a negative reward for this action.

\section{Security Service Training, Publishing and Evaluating}

\begin{figure}[ht]
    \centering
    \includegraphics[width=10.5cm]{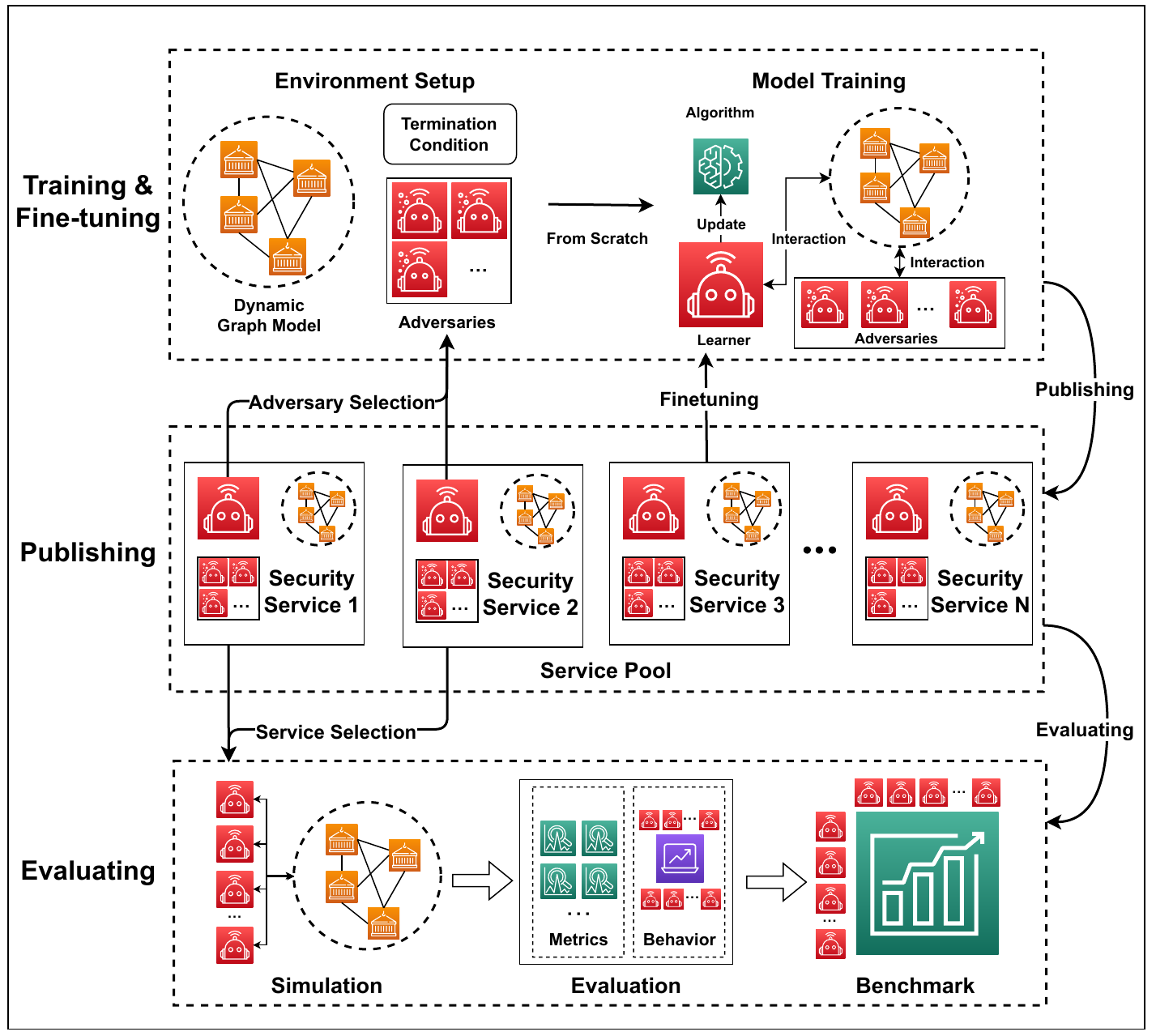}
    \caption{The Approach for Security Service Training, Publishing and Evaluating.}
    \label{fig:flow_chart}
\end{figure}

As shown in Figure \ref{fig:flow_chart}, we develop a flexible approach to train, publish and evaluate the security services. It is a three-layer framework building around the security service pool, which includes training a security service from scratch or a pre-trained one in a cloud-native environment, publishing it to the security service pool, and evaluating selected security services through simulating them in a given environment and calculate their ELO rating based performance.  

\subsection{Training a Security Service from Scratch or Pre-trained one}

The training phase aims at training an intelligent security service from scratch or from a pre-trained one for either an attacker or defender in a cloud-native environment. First, a cloud-native environment will be set up using the dynamic access graph model proposed above. This includes defining the intrinsic asset value, running state, credential connection, and the associated vulnerabilities for each node. For each associate vulnerability, we further define its attack type (local or remote), the related credential or topological information based on API endpoints, and the related gain and cost extracted from its CVSS score. These information enables us to build the access graph model for the cloud-native environment. It also sets up the termination condition, such as the maximum number of iterations and the winning goal for attackers and defenders. As we are interesting in the interactions between attackers or defenders in a cloud-native environment, rather than how attackers land on the system, we also define a specific node which would always be $owned$ by the attacker. 

The security situation depends on the interactions between attackers and defenders. Hence, when training a security service, we need to choose an adversary. More specifically, when training an offense security service, a defense security service should be selected. However, it can be $NA$, which represents the case that no defender is involved in protecting the system. On the other hand, when training a defense security service, a ${not}\ {NA}$ offense security service is required, otherwise the optimized strategy for defenders is always doing nothing. Finally, we can adopt difference advantage deep reinforcement learning algorithms, which are provided by reinforcement learning algorithm packages such as Stable-Baseline3, to train the services until reaching the termination condition. 

Furthermore, to fine-tune a pre-trained service, the same environment for the pre-trained service can be reused, or an new environment with the same action space and observation space can be set up. Then we can follow the same process described above to fine-tune the pre-trained service. 

\subsection{Publishing a Security Service to the Security Service Pool}

Once finishing training or fine-tuning, the security service can be packaged and published for further use. The published security service should include the training information, including its training environment, training algorithm and selected adversary service. Hence, we define a security service as a quintuple.

\sloppy
\textbf{Definition 10. } \textit{A Security Service is a quintuple $SS=(RO, ENV, TA, SS_{ad}, SS_{pre})$ where $RO$ is the role which can be either offense or defense, $ENV$ represents the used cloud-native environment, $TA$ represents the adopted deep reinforcement learning algorithm for training, $SS_{ad}$ is the selected adversary service for training, and $SS_{pre}$ is the selected pre-training services which can be $NA$ if trained from scratch.}

\subsection{Evaluating Security Services based on ELO Rating}

Once a security service gets published, they are available for independent performance evaluation. Note that a service's performance is related to its adversaries. Hence, we develop an ELO rating \cite{elo1978rating} based approach to evaluate a security service's related strength.


\subsubsection{Episode-based Metrics.} Given a pair of an offense security service and a defense security service $<SS_{o},SS_{d}>$ for evaluation, we will set up a simulation environment where both services can run on it and then we run the simulation to collect the following metrics:

\begin{itemize}
    \item \textbf{Average Episode Length}, $AEL$, represents the average length for the offense and defense security services from all episodes during the simulation. The lower $AEL$ the better for the attacker as a lower $AEL$ indicates the attacker takes a short time to achieve the attack goal. On the contrary, a defender would look for a higher $AEL$.
    \item \textbf{Average Episode Reward}, $AER$, represents the average reward for the offense and defense security services from all episodes during the simulation. The larger $AER$ the better for both offense and defense security services.
\end{itemize}

\subsubsection{ELO Rating based Performance.}

Initially, each service is assigned the same ELO rating score. To compare the relative strength of two offense security services $SS_{o,i}$,$SS_{o,j}$ in attacking a given cloud environment $DG_e$, we run a separate simulation for each attack service with the same defense security service $SS_{d,k}$ as the adversary, and calculate the episode-based metrics described above. If $SS_{o,i}$ has a lower $AEL$, or a higher $AER$ when $AEL$ are the same, we consider $SS_{o,i}$ wins $SS_{o,j}$ in attacking environment $DG_e$ which is defended by $SS_{d,k}$. Otherwise, $SS_{o,j}$ wins. Afterward, following the below equations, we can update the ELO rating score for each service, where $R$ stands for its ELO rating score, $E$ stands for the expected score, $S$ stands for the outcome and $S=1$ if a service wins the simulation, $S=0$ if lost and $S=0.5$ if both services have the same metrics.

\begin{equation}
E_{SS_{o,i}}=\frac{1}{1+10^{\left ( R_{SS_{o,i}}-R_{SS_{o,j}} \right )/400}}
\end{equation}
\begin{equation}
{R_{SS_{o,i}}}'= R_{SS_{o,i}}+K\times \left ( S_{SS_{o,i}} - E_{SS_{o,i}} \right )
\end{equation}

Finally, by iterating all the potential defense security services $SS_{d,k} \in SS_{d}$ and other offense security services $SS_{o,i} \in SS_{o}$ from the selected services for bench-marking, we can get an average ELO rating score for $SS_{o,i}$, representing its relative capability in attacking a given cloud environment $DG_e$. Note that this ELO rating score is comparable among different offense security services selected for benchmarking. Hence, we can rank these services and conduct a benchmark to compare their capability in attacking the given environment. Similarly, we can follow the same process to evaluate the defense security services' capability to protect a given cloud environment by calculating their ELO rating score. 
\section{Experiments and Results}
We implement our framework by customizing CyberBattleSim\footnote{https://github.com/microsoft/CyberBattleSim. Note that our framework can also be built on the other ACO Gyms such as CybORG mentioned in Section 2.2}, which provides a typical ACO Gyms-based environment, and Stable-Baseline3\footnote{https://github.com/DLR-RM/stable-baselines3}, which provides a series of advanced deep reinforcement learning algorithms. We also develop an adaptor to connect these two components loosely, making our system flexible to different ACO Gyms and reinforcement learning algorithms. 

\subsection{A Three-service-chain cloud-native System}

\begin{figure}[ht]
    \centering
    \includegraphics[width=12cm]{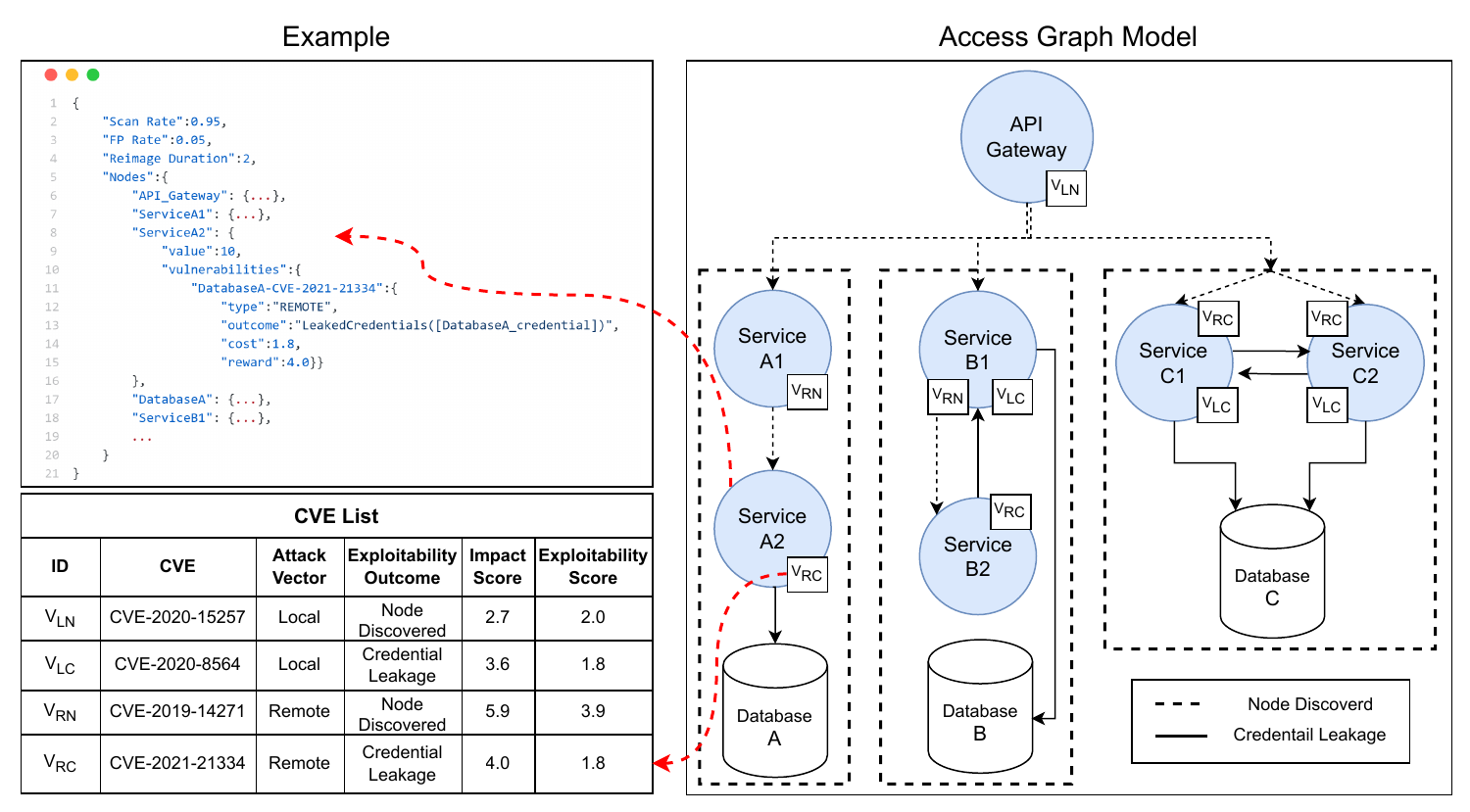}
    \caption{A three-service-chain cloud-native System. The right part is the access graph model for the environment, including the nodes, their links through API endpoint information (dot link) or access credential (solid link), and vulnerabilities associated with each node. The upper left part is the JSON-based configuration for $service A2$, provided as an example. The lower part reports the associated CVEs and their CVSS scores in the environment.}
    \label{fig:Env}
\end{figure}

Following the cloud-native system design proposed in \cite{marchese2022extending}, we propose a three-service-chain cloud-native system, as shown in Figure \ref{fig:Env} to validate our framework. It includes an API Gateway service distributing requests to three different independent service chains, which is also set as the landing point for attackers. Each service chain contains two microservices and one database server. The attacker's goal is to connect to these three databases, while the defender aims to prevent such a goal within the maximum episode length.

Additionally, each microservice could include the topological information embedded in the API endpoint, which reveals the existence of other service instances or the credential information which can be used to access others. Once a microservice is compromised by an attacker through exploiting a vulnerability, the attacker will harvest the associated topological or credential information. For example, Service $SA1$ includes the API endpoint to Service $SA2$, so the successful attack on Service $SA1$ will enable attackers to discover $SA2$. Service $SA2$ has a credential that can be used to access Database $DBA$ and compromising $SA2$ will allow attackers to $connect$ to Database $DBA$. Each microservice contains at least one vulnerability, that attackers can exploit through remote or local attack.


\subsection{Security Service Training and Publishing}

Using the above cloud-native system as the training environment, we train a series of offense and defense security services and then publish them in the security service pool. We start with training three offense security services without adversary defenders using three typical reinforcement learning algorithms, A2C, DQN, and PPO, which are provided by Stable-Baseline3. Second, using the same environment, each defense security service is trained with one offense security service as the adversary using one of the three reinforcement learning algorithms, resulting in a total of $3 \times 3 = 9$ defense security services. As reported in the upper part in Table 1, we can get 12 security services hosted in the security service pool where each security service includes its role (offense or defense), the environment, training algorithm, adversary service used for training and the selected pre-trained security services. As we aim at evaluating different security services, rather than optimizing them, we used the default parameters for each algorithm without parameter optimization and set the max episode length as 2000 and the training timesteps as 50000. Also, the $learning\_{start}$ parameter for DQN is set as 10000. 

As shown in the bottom part of Table 1, we can further fine-tune existing services using the published adversary services, which are not necessarily the same as before. For example, we can use offense security service $AA$ as the adversary to further fine-tune the three defense security services $DDA$,$DDD$, and $DDP$, which were previously trained using offense security service $AD$ as the adversary. This results in three new security services $DDAA$, $DDAD$, and $DDAP$. Similarly, we can use $AA$ as the adversary to continuously fine-tune those previously trained with $AA$, or use $AD$ as the adversary to fine-tune those previously trained with $AD$. This will result in 9 defense security services trained with different combinations of offense security services. Overall, our approach can easily train a series of security services in a systematic and flexible way. It can also incorporate complex combinations of strategies, which can reflect the fact that both attackers and defenders can adopt diverse strategies for security operations.

\begin{table}[ht]

\centering
\label{table:SPTS}
\caption{Security Service Pool with Services trained from scratch or pre-trained}
\scriptsize
\begin{tabular}{lccccc}
\Xhline{1pt}
\multicolumn{6}{l}{Environment: Three-service-chain cloud-native System} \\ \Xhline{1pt}
\multicolumn{1}{l|}{Train Type} &
  \multicolumn{1}{c|}{Service(ID)} &
  \multicolumn{1}{c|}{Role(RO)} &
  \multicolumn{1}{c|}{Algorithm(TA)} &
  \multicolumn{1}{c|}{Adversary($SS_{ad}$)} &
  Pre-Train($SS_{pre}$) \\ \Xhline{1pt}
\multicolumn{1}{l|}{\multirow{12}{*}{\begin{tabular}[c]{@{}l@{}}From\\ Scratch\end{tabular}}} &
  \multicolumn{1}{c|}{AA} &
  \multicolumn{1}{c|}{\multirow{3}{*}{Attacker}} &
  \multicolumn{1}{c|}{A2C} &
  \multicolumn{1}{c|}{\multirow{3}{*}{\textbackslash{}}} &
  \multirow{12}{*}{\textbackslash{}} \\ \cline{2-2} \cline{4-4}
\multicolumn{1}{l|}{} &
  \multicolumn{1}{c|}{AD} &
  \multicolumn{1}{c|}{} &
  \multicolumn{1}{c|}{DQN} &
  \multicolumn{1}{c|}{} &
   \\ \cline{2-2} \cline{4-4}
\multicolumn{1}{l|}{} &
  \multicolumn{1}{c|}{AP} &
  \multicolumn{1}{c|}{} &
  \multicolumn{1}{c|}{PPO} &
  \multicolumn{1}{c|}{} &
   \\ \cline{2-5}
\multicolumn{1}{l|}{} &
  \multicolumn{1}{c|}{DAA} &
  \multicolumn{1}{c|}{\multirow{9}{*}{Defender}} &
  \multicolumn{1}{c|}{A2C} &
  \multicolumn{1}{c|}{\multirow{3}{*}{AA}} &
   \\ \cline{2-2} \cline{4-4}
\multicolumn{1}{l|}{} &
  \multicolumn{1}{c|}{DAD} &
  \multicolumn{1}{c|}{} &
  \multicolumn{1}{c|}{DQN} &
  \multicolumn{1}{c|}{} &
   \\ \cline{2-2} \cline{4-4}
\multicolumn{1}{l|}{} &
  \multicolumn{1}{c|}{DAP} &
  \multicolumn{1}{c|}{} &
  \multicolumn{1}{c|}{PPO} &
  \multicolumn{1}{c|}{} &
   \\ \cline{2-2} \cline{4-5}
\multicolumn{1}{l|}{} &
  \multicolumn{1}{c|}{DDA} &
  \multicolumn{1}{c|}{} &
  \multicolumn{1}{c|}{A2C} &
  \multicolumn{1}{c|}{\multirow{3}{*}{AD}} &
   \\ \cline{2-2} \cline{4-4}
\multicolumn{1}{l|}{} &
  \multicolumn{1}{c|}{DDD} &
  \multicolumn{1}{c|}{} &
  \multicolumn{1}{c|}{DQN} &
  \multicolumn{1}{c|}{} &
   \\ \cline{2-2} \cline{4-4}
\multicolumn{1}{l|}{} &
  \multicolumn{1}{c|}{DDP} &
  \multicolumn{1}{c|}{} &
  \multicolumn{1}{c|}{PPO} &
  \multicolumn{1}{c|}{} &
   \\ \cline{2-2} \cline{4-5}
\multicolumn{1}{l|}{} &
  \multicolumn{1}{c|}{DPA} &
  \multicolumn{1}{c|}{} &
  \multicolumn{1}{c|}{A2C} &
  \multicolumn{1}{c|}{\multirow{3}{*}{AP}} &
   \\ \cline{2-2} \cline{4-4}
\multicolumn{1}{l|}{} &
  \multicolumn{1}{c|}{DPD} &
  \multicolumn{1}{c|}{} &
  \multicolumn{1}{c|}{DQN} &
  \multicolumn{1}{c|}{} &
   \\ \cline{2-2} \cline{4-4}
\multicolumn{1}{l|}{} &
  \multicolumn{1}{c|}{DPP} &
  \multicolumn{1}{c|}{} &
  \multicolumn{1}{c|}{PPO} &
  \multicolumn{1}{c|}{} &
   \\ \Xhline{1pt}
\multicolumn{1}{l|}{\multirow{9}{*}{Finetune}} &
  \multicolumn{1}{c|}{DAAA} &
  \multicolumn{1}{c|}{\multirow{9}{*}{Defender}} &
  \multicolumn{1}{c|}{A2C} &
  \multicolumn{1}{c|}{\multirow{3}{*}{AA}} &
  DAA \\ \cline{2-2} \cline{4-4} \cline{6-6} 
\multicolumn{1}{l|}{} &
  \multicolumn{1}{c|}{DAAD} &
  \multicolumn{1}{c|}{} &
  \multicolumn{1}{c|}{DQN} &
  \multicolumn{1}{c|}{} &
  DAD \\ \cline{2-2} \cline{4-4} \cline{6-6} 
\multicolumn{1}{l|}{} &
  \multicolumn{1}{c|}{DAAP} &
  \multicolumn{1}{c|}{} &
  \multicolumn{1}{c|}{PPO} &
  \multicolumn{1}{c|}{} &
  DAP \\ \cline{2-2} \cline{4-6} 
\multicolumn{1}{l|}{} &
  \multicolumn{1}{c|}{DDAA} &
  \multicolumn{1}{c|}{} &
  \multicolumn{1}{c|}{A2C} &
  \multicolumn{1}{c|}{\multirow{3}{*}{AA}} &
  DDA \\ \cline{2-2} \cline{4-4} \cline{6-6} 
\multicolumn{1}{l|}{} &
  \multicolumn{1}{c|}{DDAD} &
  \multicolumn{1}{c|}{} &
  \multicolumn{1}{c|}{DQN} &
  \multicolumn{1}{c|}{} &
  DDD \\ \cline{2-2} \cline{4-4} \cline{6-6} 
\multicolumn{1}{l|}{} &
  \multicolumn{1}{c|}{DDAP} &
  \multicolumn{1}{c|}{} &
  \multicolumn{1}{c|}{PPO} &
  \multicolumn{1}{c|}{} &
  DDP \\ \cline{2-2} \cline{4-6} 
\multicolumn{1}{l|}{} &
  \multicolumn{1}{c|}{DDDA} &
  \multicolumn{1}{c|}{} &
  \multicolumn{1}{c|}{A2C} &
  \multicolumn{1}{c|}{\multirow{3}{*}{AD}} &
  DDA \\ \cline{2-2} \cline{4-4} \cline{6-6} 
\multicolumn{1}{l|}{} &
  \multicolumn{1}{c|}{DDDD} &
  \multicolumn{1}{c|}{} &
  \multicolumn{1}{c|}{DQN} &
  \multicolumn{1}{c|}{} &
  DDD \\ \cline{2-2} \cline{4-4} \cline{6-6} 
\multicolumn{1}{l|}{} &
  \multicolumn{1}{c|}{DDDP} &
  \multicolumn{1}{c|}{} &
  \multicolumn{1}{c|}{PPO} &
  \multicolumn{1}{c|}{} &
  DDP \\ \Xhline{1pt}
\end{tabular}
\end{table}

\subsection{Security Service Evaluating using ELO Rating Score}

We can select security services hosted in the security service pool for bench-marking. At least one offense and defense security service should be selected, and the selected services should be compatible with the same environment. In other words, they should be able to simulate in the same cloud-native environment. 

Then for each pair of offense and defense security services within the selected list, we simulate their interaction using the environment designed above. For each simulation, we ran 50 episodes to calculate the Average Episode Length $AEL$ and Average Episode Reward $AER$, and then repeated 25 simulations to calculate the ELO rating score. In this research, we set the initial rating as 1000 and the K-factor as 32\footnote{We follow the existing Chess system for initial parameter setting. Please refer to this link for more details: https://www.chessclub.com/assets/Dasher/Ratings.htm.}. 

\begin{figure}[t]
    \centering
    \includegraphics[width=12cm]{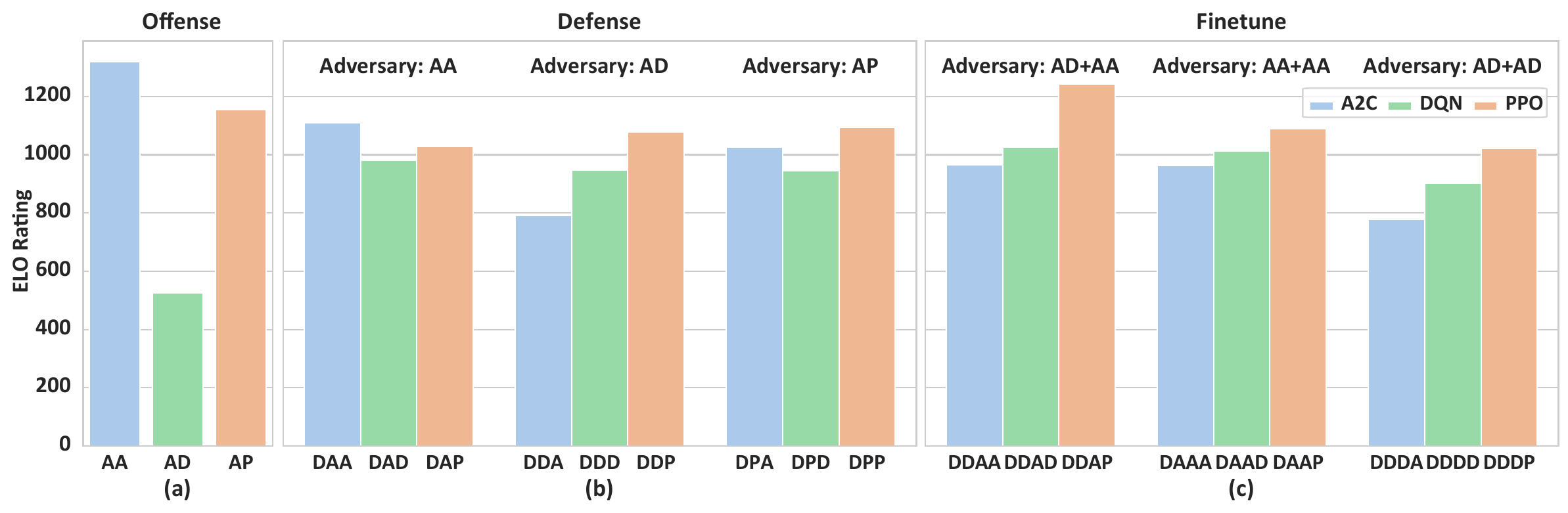}
    \caption{Performance Benchmark for Selected Security Services in Table 1. }
    \label{fig:ELO}
\end{figure}

As shown in the upper part of Figure \ref{fig:ELO} (a) and Figure \ref{fig:ELO} (b), we choose the three offense security services and night defense security services, which are trained from scratch for bench-marking. Intuitively, among the three offense security services, service $AA$, trained using $A2C$ algorithm, achieves the best performance, while service $AD$, trained using $DQN$ algorithm, has the worse performance in attacking the given cloud-native environment. For the defense security services, service $DAA$, trained using $A2C$ algorithm with offense security service $AA$ as the adversary, achieves the best performance, while service $DDA$, trained using $A2C$ algorithm with offense security service $AD$ as the adversary, achieves the worse performance. This suggests that \textit{the defense security service using a better offense security service as the adversary will have a better performance}. 

Finally, we are interested in the impact of the fine-tuning strategy. As shown in Figure \ref{fig:ELO} (c), the defense security service trained based on two different offense security services always outperforms the defense security service trained based on the same offense service. While this confirms our framework's capability in training services using complex strategies, it suggests that \textit{when training a defense security service, using different offense security services as the adversaries for fine-tuning can improve the defense security service's capability}. 

\section{Conclusion}

Using the multi-agent deep reinforcement learning paradigm, we develop the agent-based intelligent security service framework (ISSF) for cloud-native security operations. It includes a dynamic access graph model representing the cloud-native environment from a defense aspect, the action space model representing the attackers' and defenders' actions, and a flexible approach to train, publish and evaluate the offense and defense security services. The experiment on a three-service-chain system confirms that our framework can effectively develop and quantitatively evaluate diverse, from simple to complex, intelligent security services for attackers and defenders in the given cloud-native environment. 

While the preliminary result confirms the effectiveness of our framework, our observations also open a gateway for future strategies optimization, such as incorporating advanced and diverse adversaries for training and fine-tuning.  Additionally, further studies to investigate different parameters' influence on the stability of the ELO ranking are also valuable. Finally, while the current performance evaluation is based on a simulated cloud-native environment, extending the evaluation on the emulated system would also be essential.

\bibliographystyle{splncs04}

%





\newpage

\section{Support Material}

\begin{table}[]
\centering
\begin{tabular}{l|l|l|l|l|l|l}
\hline
Attack Type\textbackslash{}Tactics &
  \begin{tabular}[c]{@{}l@{}}Initial\\ Access\end{tabular} &
  Execution &
  Persistence &
  \begin{tabular}[c]{@{}l@{}}Privilege\\ Escalation\end{tabular} &
  \begin{tabular}[c]{@{}l@{}}Defense\\ Evasion\end{tabular} &
  \begin{tabular}[c]{@{}l@{}}Credential\\ Access\end{tabular} \\ \hline
Local   & \ding{51} & \ding{51} & \ding{51} & \ding{51} & \ding{51} & \ding{51} \\ \hline
Remote  & \ding{51} & \ding{51} & \ding{51} & \ding{51} & \ding{51} & \ding{51} \\ \hline
Connect & \ding{51} &   &   & \ding{51} &   &   \\ \hline
Attack Type\textbackslash{}Tactics &
  Discovery &
  \begin{tabular}[c]{@{}l@{}}Lateral\\ Movement\end{tabular} &
  Collection &
  Exfiltration &
  Impact &
   \\ \hline
Local   & \ding{51} &   & \ding{51} & \ding{51} & \ding{51} &   \\ \hline
Remote  & \ding{51} &   & \ding{51} & \ding{51} & \ding{51} &   \\ \hline
Connect &   & \ding{51} &   &   &   &   \\ \hline
\end{tabular}
\end{table}

\end{document}